\documentclass[prb,twocolumn, superscriptaddress]{revtex4}
\usepackage{graphicx}
\begin{document}
\title{Measuring anisotropic scattering in the cuprates}
\author{K. G. Sandeman}
\affiliation{Low Temperature Physics Group, Cavendish Laboratory,
Madingley Road, Cambridge, CB3 0HE, United Kingdom.}
\author{A. J. Schofield}
\affiliation{School of Physics and Astronomy, University of
Birmingham, Edgbaston, Birmingham B15 2TT, United Kingdom.}
\date{\today}
\begin{abstract}
A simple model of anisotropic scattering in a quasi two-dimensional
metal is studied. Its simplicity allows an analytic calculation of
transport properties using the Boltzmann equation and relaxation time
approximation. We argue that the $c$-axis magnetoresistance provides
the key test of this model of transport.  We compare this model with
experiments on overdoped Tl-2201 and find reasonable agreement using
only weak scattering anisotropy. We argue that optimally doped Tl-2201
should show strong angular-dependent magnetoresistance within this
model and would provide a robust way of determining the in-plane
scattering anisotropy in the cuprates.
\end{abstract}
\maketitle


\section{Introduction} 
The unusual transport properties in the normal state of the cuprates
continue to attract widespread interest. Many aspects differ
significantly
from transport in conventional metals~\cite{iye_1992a}.  The
resistivity is linear in temperature~\cite{gurvitch_1987a} with
scattering which is apparently electronic rather than electron-phonon
in origin~\cite{bonn_1993a}.  The Hall effect shows strong temperature
dependence~\cite{iye_1992a}.  On doping with zinc, both the
resistivity and inverse Hall angle show a rather simple Matthiessen's
rule~\cite{chien_1991a}, suggestive of additive scattering
mechanisms. However, their temperature dependences differ
significantly.  This has led to the suggestion that the resistivity
and the inverse Hall angle are controlled by very different scattering
rates (the so-called `two-lifetime'
behavior)~\cite{anderson_1991a}. This view seems to be confirmed by
optical measurements~\cite{kaplan_1996a} which suggest two independent
scattering rates in these materials.  The magnetoresistance does not
follow Kohler scaling behavior but appears also to be controlled by
the Hall angle scattering rate~\cite{harris_1995a}.  Attempts to
understand this behavior may be divided into three categories.

Anderson~\cite{anderson_1991a} has argued that a non-Fermi liquid
description must be invoked to understand transport in the
cuprates. In Anderson's picture the electron decays into holons and
spinons which separately control resistivity and magneto-transport
respectively. A phenomenological transport equation was introduced by
Coleman, Schofield and Tsvelik~\cite{schofield_1996a} to capture a
model of transport where the two lifetimes are controlled by
independent fluids of particles.

A second approach argues that the magnetic field plays a special role
in the transport process. In the model of Kotliar, Sengupta and
Varma~\cite{kotliar_1996a} a singular skew scattering term is
invoked. In the picture by Lee and Lee~\cite{lee_1997a} only
recombined slave-fermion and slave-boson particles can interact with
the true applied magnetic field. The slave particles themselves are
insensitive to the true field because of the large fluctuations in the
fictitious gauge field. In both of these scenarios the measured
cyclotron frequency would appear to be temperature dependent---a
result at odds with current optical Hall
experiments~\cite{drew_1999a}.

The third possibility---the `anisotropic scattering'
scenario---envisages a conventional metallic state characterized by an
electron quasiparticle scattering rate which depends strongly on the
particle momentum.  This picture was first suggested by
Cooper~\cite{carrington_1992a} and has been further elaborated upon by
Stojkovi\'c and Pines~\cite{stojkovic_1996a}, and also by Yakovenko
and coworkers~\cite{yakovenko_1998a,yakovenko_1999a}. Pines {\it et
al.} argue that antiferromagnetic fluctuations in the normal
state strongly scatter regions on the Fermi surface which are
connected by the antiferromagnetic wave vector.  Regions away from
these `hot-spots' are weakly scattered. However, Hlubina and
Rice~\cite{hlubina_1995a} have claimed that the cold regions would
tend to `short-circuit' the hot spots, thereby leading to conventional
transport at low temperatures. Ioffe and Millis~\cite{ioffe_1998a}
have introduced a new variant on the anisotropic scattering picture
which is partly motivated by the short-circuiting problem and also
inspired by the direct measurements of electron lifetimes.  They use
angle-resolved photoemission data to argue that quasiparticles are
very strongly scattered over most of the Fermi surface.  The only
long-lived quasiparticles are found at the zone diagonals, where they
decay with a weak $T^2$ Fermi liquid form.  This is known as the `cold
spot' model and has been used to understand a wide variety of
experiments (see for example Van der Marel~\cite{vanderMarel_1999a} and
Xiang and Hardy~\cite{xiang_2000a}).

In this paper we consider this third scenario of anisotropic
scattering and treat a minimal model which we believe is
illustrative of the strengths and weaknesses of this approach
(similar in spirit to some treatments of the cuprate superconducting
state~\cite{maki_1995a}).  Whilst more sophisticated treatments
exist, either with a specific relaxation
mechanism~\cite{stojkovic_1996a,rosch_1999a} or in the limit of
extremely anisotropic in-plane scattering~\cite{ioffe_1998a}, there
are advantages to this model's simplicity.  We can obtain analytic
expressions for a wide range of transport properties over the full
range of anisotropy. Our calculations illustrate the conclusions of
Ioffe and Millis in the limit of strong anisotropy, but also allow
us to see how this evolves from the more familiar isotropic metal.  
This is important as the approach to the isotropic limit may be more
relevant to experiments on overdoped materials where the
two-lifetime behavior is less apparent.  We also consider the
transition to the intermediate and high field regimes of
magnetoresistance within this picture, and non-linear effects at
high electric field.

Perhaps the key discrepancy between the `cold-spot' model and the
transport properties of the cuprates is the in-plane orbital
magnetoresistance. The model predicts a large magnetoresistance with a
distinct temperature dependence in contrast with current experiments.
This was was noted in Ioffe and Millis' original paper~\cite{ioffe_1998a} 
and is reproduced
in passing here. It has been argued that proper inclusion of vertex
corrections could account for this
discrepancy~\cite{narikiyo_2000a,kontani_1999a}. The most important
new work in this paper is to consider the orbital magnetoresistance
for currents moving along the $c$-axis. The $c$-axis conductivity in a
quasi-two dimensional metal may be computed {\em without} vertex
corrections~\cite{millis_2000a} and so should be a robust feature of
this scenario. We show how an in-plane magnetic field affects the
$c$-axis conductivity and argue that this experimental geometry
provides a key test of the model. Indeed experiments exist on
Tl-2201~\cite{hussey_1996a} but only in the overdoped regime where the
two lifetimes become less distinct. Nevertheless our complete solution
of an anisotropic scattering model allows us to obtain a reasonable
fit to the data provided we include both scattering anisotropy and
bandstructure effects.

Our main conclusions are that within our model the in-plane
magnetoresistance remains an outstanding problem if this model is to
fit the experiment---not just from its magnitude and temperature
dependence but from the scale at which deviations from the weak-field
regime occur. While a proper treatment of vertex corrections may
correct this for in-plane properties, the $c$-axis magnetoresistance
is insensitive to these corrections and provides a robust test of the
model. Experiments here have mainly been done on overdoped cuprates
and our detailed fit to experiments are consistent with this
model. High field $c$-axis measurements on optimally doped cuprates
should be made to test unambiguously this transport phenomenology. 

This paper is planned as follows. In section~\ref{Aniso} we present our
models of scattering and bandstructure which are the ingredients for the
analytic calculation of transport properties.  The results are to be found
in Section~\ref{Results}.  These are displayed in the general case, and
also in the limit of large scattering anisotropy.  Both in-plane and
out-of-plane magnetoresistivity are shown, and compared with experimental
observations in section \ref{application}.  Non-ohmic in-plane
conductivity is also calculated. Conclusions are drawn in section
\ref{Conclusion}.

\section{Anisotropic transport}
\label{Aniso}

Our main assumption in this work is to use a Boltzmann equation within
the relaxation time approximation as a description of transport.  We
write the collision integral, $I[f]$, as $-\Gamma_{\mathbf
k}[f({\mathbf k})-f^{(0)}({\mathbf k})]$ and impose a phenomenological
scattering rate, $\Gamma_{{\mathbf k}}$, which varies smoothly around
the in-plane Fermi surface and has square symmetry.

We do not speculate on the underlying cause of the quasiparticle
relaxation rate but, if this scattering were due to scattering from a
soft bosonic mode at finite ${\mathbf Q}$ (antiferromagnetic spin
fluctuations for example), then the validity of the relaxation time
approximation might be questionable~\cite{rosch_2000a}.  This is
because such scattering would tend to equilibrate quasiparticles only
between points on the Fermi surface connected by the spanning vector
(${\mathbf Q}$) of the mode. By contrast, the relaxation time
approximation assumes that the quasiparticle is equilibrated
uniformly around the Fermi surface.  This will not affect our
calculation of c-axis response if the bosonic mode is 2-dimensional,
but it could change the in-plane physics.  This is presumably the
reason why vertex corrections could be important for in-plane
transport.  However, including this process correctly would tend to
make the quasiparticle distribution even more anisotropic when driven
out of equilibrium. As we will see, the anisotropy of the
out-of-equilibrium distribution of quasiparticles is precisely what is
probed by the in-plane magnetoresistance. Increasing this anisotropy
is likely to increase further the in-plane magnetoresistance.

Within the relaxation time approximation we now introduce two forms of
scattering rate to model the anisotropy. 

\subsection{The cold-spot model}
\label{Modelcold}
Following Ioffe and Millis~\cite{ioffe_1998a}, the angle-resolved
photoemission results suggest that there are points on the Fermi
surface where the electron is a fairly long-lived excitation: points
where the Fermi surface intersects the Brillouin zone
diagonal. Elsewhere on the Fermi surface, only broad features in
energy are seen in the photoemission.  Within a Fermi liquid picture
one would then infer that electron-like quasiparticles are strongly
scattered everywhere except for these special points. We can capture 
this by the following form for our scattering rate
\begin{equation}  
\Gamma(k_{\parallel},\theta,k_{\perp})=\Gamma_{f}\cos ^2 2\theta 
+\Gamma_{s}\sin ^2 2\theta \, .
\label{scatrate1}
\label{coldrate}  
\end{equation}

\begin{figure}
\includegraphics[width=\columnwidth]{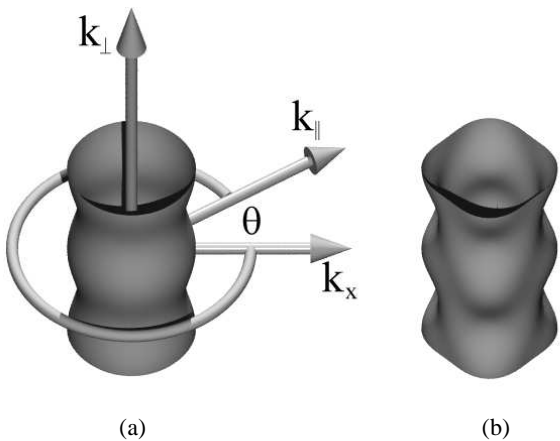}
\caption{The cylindrical coordinates used in this paper.  
This also illustrates the Fermi surface we consider with the $c$-axis
dispersion greatly exaggerated. The degree to which the $c$-axis
dispersion depends on the in-plane momentum varies within this model
from (a) no dependence ($\gamma=0$) to (b) the case when there is no
dispersion along the zone diagonals ($\gamma=1$).} 
\label{figure1} 
\end{figure}

\begin{figure}
\includegraphics[width=\columnwidth]{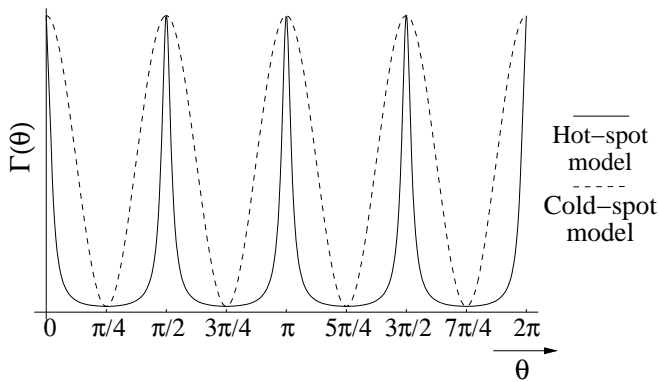} 
\caption{The angular dependence of the scattering rate in our
parameterization of the cold-spot model (dotted line and
Eq.~\ref{coldrate}) and the hot-spot model (solid line and
Eq.~\ref{hotrate}). Although the current is preferentially carried by
the longest-lived quasiparticle, in the cold-spot model the
short-lived particles can contribute since they cover most of the
Fermi line. In the hot-spot model the cold regions short circuit all
transport processes and dominate all quantities.}
\label{figure2}
\end{figure}

Here $\theta$ is the azimuthal angle between the in-plane momentum wavevector 
$k_{\parallel}$ and the $a$-axis direction of the crystal, as shown in
Fig.~\ref{figure1}.
%
This is a simple model which nevertheless allows us to
explore the transition from isotropic ($\Gamma_f=\Gamma_s$) to
extremely anisotropic scattering ($\Gamma_f \gg \Gamma_s$).  The
latter limit corresponds to the case studied by Ioffe and Millis,
where, at the planar zone diagonals [$\theta=(2n +1)\pi/4$] the
scattering rate has a minimum.  In their model, $\Gamma_s$ is just the
Fermi-liquid scattering rate, $\Gamma_s = 1/\tau_{FL}\sim T^2$. The
scattering around the Fermi-surface in the limit of strong anisotropy
is illustrated in Fig.~\ref{figure2}. 
%

It will often be convenient to express the scattering rate in terms of
an anisotropy parameter, $\alpha$, where Eq.~\ref{scatrate1} is written as 
\begin{equation}  
\Gamma(k_{\parallel},\theta,k_{\perp}) =
\Gamma_{0}\bigl(1+\alpha\cos{4\theta}\bigr)
\; ,
\label{scatrate}  
\end{equation}
with
\begin{eqnarray}  
\Gamma_{0} &=&(\Gamma_{f}+\Gamma_{s})/2 \; ,\\
\alpha &=& {(\Gamma_{f}-\Gamma_{s}) \over (\Gamma_{f}+\Gamma_{s})} \; .
\label{scatdefn}
\end{eqnarray}
In this form $\alpha=0$ gives us the isotropic scattering case whilst
$\alpha=1$ or $-1$ gives large anisotropy with cold spots on the zone
diagonals or zone axes respectively.  We will express results in either
notation ($\Gamma_f, \Gamma_s$ or $\alpha, \Gamma_0$) depending on
applicability to available data and ease of interpretation, and are
always able to calculate across the full range of $\alpha$.

\subsection{The hot-spot model}
\label{Modelhot}
Alternatively it is has been proposed that 
hot spots exist near the $(\pi,0)$ points of the Brillouin zone
where strong scattering occurs, perhaps due to antiferromagnetic spin
fluctuations. This leaves quasiparticles on the rest of the 
Fermi surface with a longer lifetime. To capture this within our
phenomenology we add the scattering lifetimes around the Fermi surface so 
\begin{equation}
\Gamma^{-1}_{\mathbf k} =
\tau(k_{\parallel},\theta,k_{\perp})=\tau_{f}\cos ^2 2\theta +\tau_{s}\sin
^2 2\theta \; . 
\label{hotrate} 
\end{equation} 
Here we expect $\tau_f$ to be smaller than $\tau_s$ so that  
we have a model with strong-scattering hot spots and extended cold
regions with less scattering (see Fig.~\ref{figure2}).

\subsection{Bandstructure}
\label{bandstructure}

\begin{figure}
\includegraphics[width=\columnwidth]{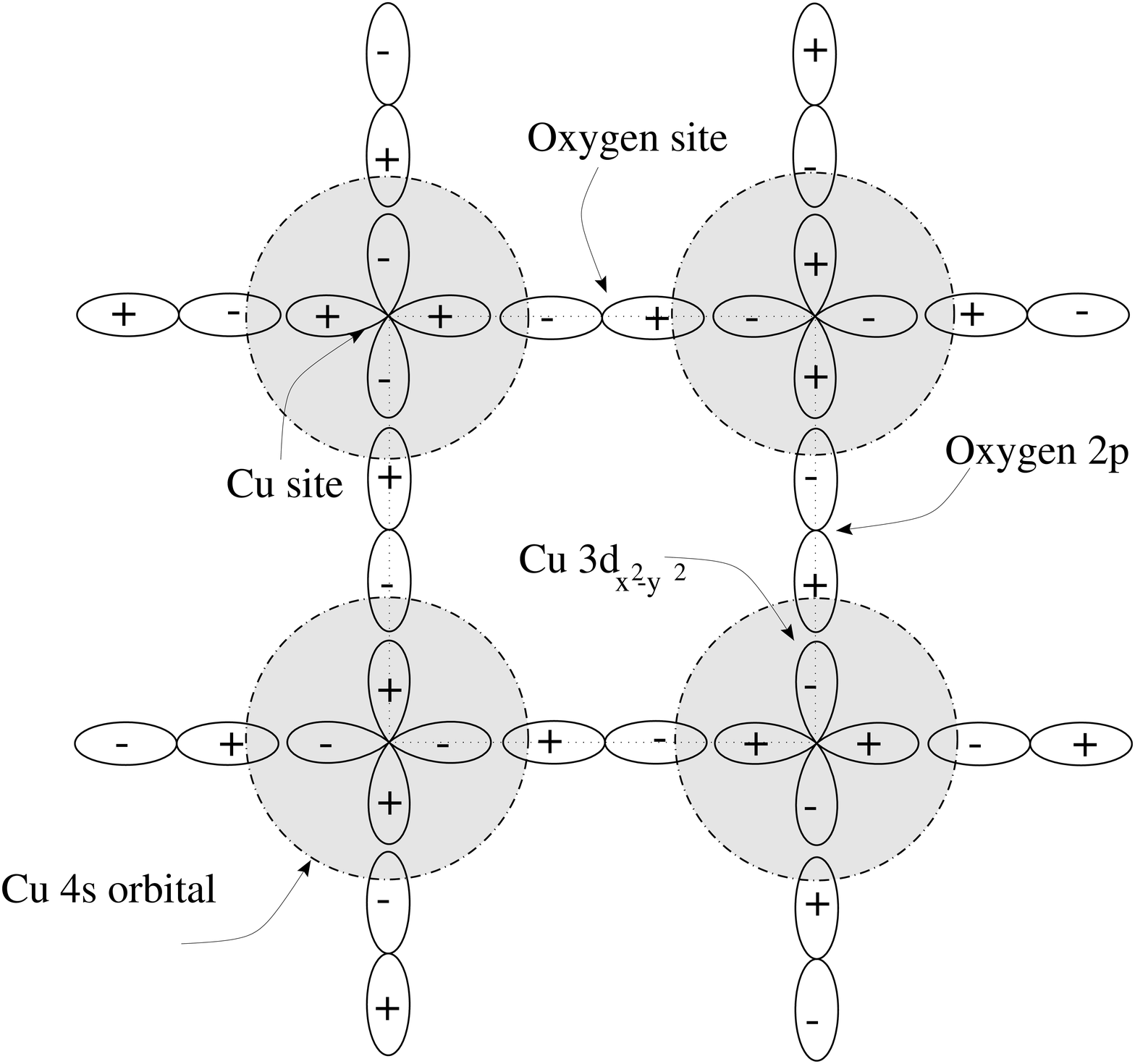} 
\caption{The overlap between the hybrids of copper
3d$_{x^2-y^2}$ and oxygen 2p orbitals and a copper 4s has d-wave
symmetry. Since transport between layers occurs primarily via the
copper 4s
orbitals, the $c$-axis tunneling probability vanishes when the
in-plane quasiparticle momentum is directed along the zone
diagonals---the zeros of the $d_{x^2-y^2}$ form (after Xiang {\it et
al.}~\cite{xiang_1998a}).}
\label{figure3}
\end{figure}

The next component of the model is the bandstructure. We assume
that the in-plane dispersion is isotropic, which is not unreasonable
for the cuprates. By contrast the dispersion in the out-of-plane
direction is known to have significant dependence on the in-plane
momentum~\cite{xiang_1996a,andersen_1994a,liechtenstein_1996a}.  This
is due to the local chemistry of the copper-oxide planes as emphasized
by Xiang~\cite{xiang_1998a}. Transport between planes
occurs principally through the copper 4s orbitals.  Since these
orbitals are circularly symmetric in the plane, their overlap
amplitude with the
Cu 3d$_{x^2-y^2}$--O 2p hybrids has d-wave symmetry as illustrated in
Fig.~\ref{figure3}.  
The resulting tunneling term is therefore
proportional to the square of this amplitude and can be represented by
the form $t_{\perp}\cos^2{(2\theta)}$. The most important effect this
has on $c$-axis transport stems from the fact that the longest-lived
quasiparticles on the zone diagonals are precisely those that have a
vanishing probability of $c$-axis hopping. For this reason we ignore
other features of the $c$-axis dispersion which may lead to additional
small tunneling probabilities at other points on the Fermi-surface
such as those due to a body-centered tetragonal unit cell considered
elsewhere~\cite{dragulescu_1999a,vanderMarel_1999a}. Our
bandstructure is then given by

%
%
%
\begin{equation}  
\epsilon({\mathbf k_{\parallel}},\theta,{\mathbf k_{\perp}})=
\epsilon(k_{\parallel}^{2})
-2t_{\perp}\bigl(1+\gamma
\cos{4\theta}\bigr)\cos(k_{\perp}c) \, .
\label{dispersion}  
\end{equation}
To illustrate the importance of this effect and to allow for other
tunneling mechanisms we parameterize the degree of tunneling
anisotropy by $\gamma$, which we allow to vary from 0 to 1.  When
$\gamma$=1, we have complete elimination
of c-axis transport for quasiparticles with in-plane momentum along
the zone diagonals.  The frequency dependence of this end-point of the
model has been studied by van der Marel~\cite{vanderMarel_1999a}.
Allowing both the degree of scattering
anisotropy ($\alpha$) and $c$-axis dispersion ($\gamma$) to vary across
their full range turns out to be crucial in understanding
the out-of-plane transport and is a novel aspect of this work. It unifies
various other
approaches~\cite{stojkovic_1996a,ioffe_1998a,hussey_1996a} which focus
on the extremes of one or other of the ranges of $\alpha$ and
$\gamma$.
%
%
%
\section{Results}
\label{Results}
Using the standard representation of the quasiparticle distribution,
$f = f^{(0)} + \psi\partial_\epsilon f^{(0)}$, we
write the Boltzmann equation (for uniform fields and temperature
gradients) as~\cite{abrikosov_1980a}
\begin{eqnarray}
-\partial_t \psi + \Gamma_{{\mathbf k}}\psi- {e \over
\hbar}({\mathbf E} + {\mathbf v_{\mathbf k}} \times
{\mathbf B})\cdot{\nabla_{{\mathbf k}}\psi} &=& \nonumber \\
e{\mathbf E}\cdot{\mathbf
v_{\mathbf k}} &-& {\epsilon_{\mathbf k} -\mu \over T} 
{\mathbf v_{\mathbf k}} \cdot
\nabla T .
\label{simpleBoltz}
\end{eqnarray}
We will solve this equation in various limits in the subsequent
sections. We will only be interested in d.c. properties in this paper,
however our results for the cold-spot model (Eq.~\ref{coldrate}) 
are easily extended into the frequency domain.
After Fourier
transforming with respect to time, we see that 
\begin{eqnarray}
- \partial_t \psi + \Gamma(\theta) \psi & \rightarrow &
i \omega \psi + (\Gamma_f \cos^2 2\theta + \Gamma_s \sin^2 2\theta)
 \; , \nonumber \\
&\rightarrow& ( \Gamma_f + i\omega) \sin^2 2\theta + (\Gamma_s +
i\omega) \cos^2 2\theta \; . \nonumber \\
\end{eqnarray}
We can therefore obtain the finite frequency
results for the cold-spot model 
from our expressions at d.c. simply by the prescription
\begin{equation}
\Gamma_f \rightarrow \Gamma_f+i\omega \;,  \quad \quad
\Gamma_s \rightarrow \Gamma_s+i\omega \;.
\end{equation}
\subsection{In-plane transport in weak magnetic fields}
We first focus on leading order response in 
an in-plane electric field. 
For a circular Fermi surface we may write Eq.~\ref{simpleBoltz} as
\begin{equation}
\Gamma(\theta)\psi- \omega_c
{\partial \psi \over \partial \theta}=eEv_F\cos\theta \; ,
\label{simpleBoltz2}
\end{equation}
where $\omega_c=eBv_F/(\hbar k_F)$ and $v_F$ is the in-plane 
Fermi velocity.  

We may solve this equation by Jones-Zener expansion for in-plane
currents with a weak magnetic field along the $c$-axis.  The currents,
and hence conductivity elements are calculated using the relation
\begin{equation}
{\mathbf j_{\mu}}=
{eE \over 4\pi^3 \hbar}
\int\int {{{\mathbf v_{\mu}} k_{F} \over v_{F}} \psi d\theta
dk_{\perp}} \; ,
\label{currentintgl}
\end{equation}
where current is flowing in direction $\mu$ with velocity ${\mathbf
v_{\mu}}$.  We expand $\sigma_{xx}$ and $\sigma_{xy}$ to order $B^2$ and the
resistivity is then found by simple matrix inversion.

If we consider first hot-spot scattering (Eq.~\ref{hotrate}), we can
see that this scattering rate can be eliminated as a viable model for
the cuprates.  Each of the properties evaluated is dominated by the
long scattering time, which is the short-circuiting effect proposed by
Hlubina and Rice~\cite{hlubina_1995a}.  The in-plane conductivities
are given by
\begin{eqnarray}
\sigma_{xx}^{(0)}&=&{e^2 k_F v_F(\tau_f+\tau_s) \over 4\pi\hbar c} \; ,
\label{insigxx0hot} \\
{\sigma_{xy} \over
\sigma_{xx}^{(0)}} &=& \tan\Theta_{H}=
\omega_{c}{(3\tau_{f}^{2}+2\tau_{f}\tau_{s}+3\tau_{s}^{2})
\over 4(\tau_{f}+\tau_{s})} + O[B^3] \; , \nonumber \\
\label{insigxyhot} \\
{\Delta\sigma_{xx} \over \sigma_{xx}^{(0)}}&=&-\omega_{c}^2 {(21
\tau_{f}^{2}-34\tau_{f}\tau_{s}+21\tau_{s}^{2}) \over 8} +O[B^4] \; .
\label{insigxxB2hot}
\end{eqnarray}
Within this model, all computed quantities are dominated by $\tau_s$
in the limit of strong anisotropy ($\tau_s \gg \tau_f$). This model is
unable to reproduce the two-lifetime phenomenology of the cuprates so
we will not consider it further.

However, for the cold spot model we obtain the following in terms of
$\Gamma_{f}$ and $\Gamma_{s}$:
\begin{eqnarray}
\sigma_{xx}^{(0)}&=&
{e^2 v_F k_F\over 2\pi\hbar c}{1 \over \sqrt{\Gamma_{f}\Gamma_{s}}} \;
,  \label{insigxx0cold} \\
{\sigma_{xy} \over \sigma_{xx}^{(0)}}&=&
\tan\Theta_{H} = {\omega_{c}\over 2}\biggl({1 \over \Gamma_{f}} + {1
\over \Gamma_{s}}\biggr) + O[B^3] \; , 
\label{insigxycold} \\
{\Delta\sigma_{xx} \over \sigma_{xx}^{(0)}} &=&
\omega_{c}^2 \biggl[{5 \over 8}\biggl( {\Gamma_{f} \over \Gamma_{s}^3} +
{\Gamma_{s} \over \Gamma_{f}^3}\biggr) -{1 \over 8}\biggl({1 \over
\Gamma_{f}^2} +  {1 \over\Gamma_{s}^2}\biggr)\biggr] + O[B^4]\, . 
\nonumber \\
\label{insigxxB2cold}
\end{eqnarray}
This results in an in-plane orbital weak-field magnetoresistance
\begin{equation}
{\Delta\rho_{xx} \over \rho^{(0)}_{xx}} = 
{\omega_{c}^2} {(\Gamma_{f}-\Gamma_{s})^2
\over
8\Gamma_{f}^{3}\Gamma_{s}^{3}}
(5\Gamma_{f}^2 +7\Gamma_{f}\Gamma_{s} + 5\Gamma_{s}^{2}) \; ,
\label{inplanemr}
\end{equation}
which  vanishes  in   the  isotropic  limit  ($\Gamma_s=\Gamma_f$). This
reflects the well known result that there is no orbital magnetoresistance
in an isotropic metal.
In the high anisotropy ($\alpha \rightarrow 1$) limit, where
$\Gamma_f  \gg   \Gamma_s$, we find the results shown in table 
\ref{sigmatable}.
%
%
\begin{table}
\begin{tabular}{|c|ccc|}
\hline
Quantity    &  Cold-spot model   &   Experiment & Reference  \\ 
\hline
$\rho_{xx}^{(0)}$   & $\sqrt{\Gamma_f\Gamma_s}$  &    $T$ &
\onlinecite{gurvitch_1987a} \\ 
$\cot\Theta_H$     & ${2 \Gamma_s \over \omega_{c}}$  &       $T^2$ &
\onlinecite{chien_1991a}\\
${\Delta\rho_{xx} \over \rho_{xx}^{(0)}}$   &  
$\omega_{c}^{2}{\Gamma_f \over \Gamma_s^3}$ &  $\Theta_H^2$ &
\onlinecite{harris_1995a} \\
${1 \over \Theta_{H}^{2}}{\Delta\rho_{xx} 
\over \rho_{xx}^{(0)}}$                   & 
${\Gamma_f \over \Gamma_s}$ & const & \onlinecite{harris_1995a} \\
\hline
\end{tabular}
\caption{A comparison of our results with experiment in the limit
of highly-anisotropic scattering.} 
\label{sigmatable} 
\end{table}

These results illustrate the way in which the cold-spot model
reproduces the phenomenology of the cuprates as shown by Ioffe and
Millis~\cite{ioffe_1998a}. They take $\Gamma_f$ to be large and
temperature independent, while $\Gamma_s$ is assumed small and
proportional to $T^2$. The geometric mean then gives the linear in
temperature resistivity while the inverse Hall angle is proportional
to $\Gamma_s$ and hence varies as $T^2$. The results are also
illustrative of the problems of this model. Adding Zn impurities to
the ${\rm CuO_2}$ has been interpreted as adding a unitary scatterer
which should add a temperature independent term to the scattering
rates.  Adding a constant elastic scattering term to the two rates
appearing in the geometric mean which forms the resistivity
(Eq.~\ref{insigxx0cold}) will not reproduce the Matthiessen's rule
behavior seen in experiment~\cite{chien_1991a}. Instead it will
change both the temperature dependence and the intercept of the
resistivity.

The second difficulty is the large magnetoresistance that this model
predicts. While experiments suggest that the magnetoresistance is
proportional to the Hall angle squared~\cite{harris_1995a} with a constant of
proportionality which can be of order 1, the cold-spot model predicts
that these two quantities differ by $\Gamma_f/\Gamma_s$---a large,
temperature dependent factor. It is tempting to appeal to new physical
processes to explain this (such as vertex corrections). However the
cause of this large magnetoresistance is intimately related to the
appearance of distinct temperature dependences of the resistivity and
inverse Hall angle. Magnetic fields couple to the anisotropy of $\psi$
around the Fermi surface as can be seen from Eq.~\ref{simpleBoltz2}
and so at each order in the magnetic field, the effect of the
anisotropy in the model is amplified. This allows the inverse Hall
angle to differ from the resistivity but consequently has an even more
dramatic effect on the magnetoresistance which is higher order in
magnetic field. We do not attempt to offer a solution for this puzzle
though given the systematics in transport behavior across a wide range
of cuprates it seems unlikely that this is due to fine tuning. (By
contrast in the charge-conjugation
phenomenology~\cite{schofield_1996a} the magnetoresistance is
proportional to the square of the Hall angle and is small since this
model is isotropic around the Fermi surface.)

%
%
\subsection{Beyond the weak-field regime}

Thus far, in keeping with other workers, we have shown that in-plane
transport is governed by a different scattering rate (formed from the
combination of $\Gamma_f$ and $\Gamma_s$) at each order in the magnetic
field.  Here we go beyond the weak-field regime and show numerically that
no new combinations emerge beyond second order in the magnetic field until
the magnetoresistance saturates. The scattering rate combination for
weak-field magnetoresistance provides the scale for magnetic field effects
throughout the intermediate field region.

\begin{figure}
\includegraphics[width=\columnwidth]{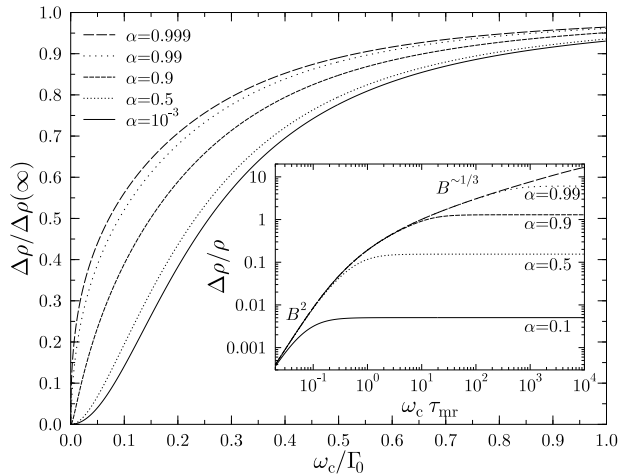} 
\caption{The in-plane magnetoresistance is shown beyond the
weak field region as determined from a numerical solution. 
Saturation occurs at a scale determined by
$\omega_c \sim 
\Gamma_0$ since quasiparticles must precess
around the whole Fermi surface. The inset shows that no new field scale
emerges between the weak-field and the saturated regime. By scaling
the data to the weak-field region (thereby defining $\tau_{mr}$) 
we see that the magnetoresistance
continues to fall on a universal curve beyond the $B^2$ regime until
the saturated magnetoresistance value is reached.}
\label{figure4} 
\end{figure}

The saturation field may be obtained from a high field expansion of
the Boltzmann equation---expanding the solution in powers of
$1/\omega_c$. We find that the magnetoresistance saturates at a value
of
\begin{equation}
{\Delta \rho_{xx}(B\rightarrow \infty) \over \rho_{xx}(B=0)} = 
{{\left(\sqrt{\Gamma_f} - \sqrt{\Gamma_s}\right)^2 \over 2
\sqrt{\Gamma_s \Gamma_f}}} \; .
\label{satscale}
\end{equation}
The high field limit is reached when $\omega_c \sim \Gamma_f$ so the
fastest rate controls the transition to the high field limit. This is
because saturation requires that quasiparticles can completely orbit
the Fermi surface and it is the hot regions which limit this
process\cite{tyler_1998a}. This is also the scale for Landau
quantization. The approach to the saturated value is shown in
Fig.~\ref{figure4}. The magnetic field is plotted as a function of
$\omega_c/\Gamma_0$ since in the cold spot model this is dominated by
$\Gamma_f$ and so shows that this sets the scale for saturation.

[As an aside, we can obtain the universal form of the magnetoresistance
in an isotropic metal. Of course, this is formally zero but by
dividing by the saturated value in the limit of small anisotropy we
can obtain a finite result. This is also shown in Fig.~\ref{figure4}.]

While we cannot solve Eq.~\ref{simpleBoltz2} analytically except in the
low and high field limits, the equation is readily solved numerically. The
fact that we require a solution of $\psi$ which is periodic around the
Fermi surface means that we can look for a Fourier series
solution. Our numerical results reproduce the analytical results of
the previous section. In
addition, we find the remarkable result that the magnetoresistance is
a function of a single dimensionless parameter until it saturates. This 
scaling parameter governs the magnetoresistance even beyond the
weak-field ($B^2$) regime of the magnetoresistance so we may write
\begin{equation}
{\Delta \rho_{xx}(B) \over \rho_{xx}(0)} = 
f(\omega_c \tau_{\rm mr}) \; , 
\quad \quad \omega_c \ll \Gamma_f \; ,
\end{equation}
where $\tau_{\rm mr}$ is defined by the weak-field magnetoresistance such
that $\Delta \rho_{xx}/\rho_{xx} = (\omega_c \tau_{\rm mr})^2$. For
$\Gamma_f \gg \Gamma_s$ we have $\tau_{\rm mr}^2 \sim
5 \Gamma_f/ 8\Gamma_s^3$.

We illustrate this universality in the inset to Fig.~\ref{figure4}. We
see that as a function of this combination of scattering rates, the
magnetoresistance follows a universal curve until the saturation
value is approached. An intermediate field region emerges for fields
$1 < \omega_c \tau_{\rm mr} < \omega_c/\Gamma_f$, where the
magnetoresistance adopts a new power law $\sim B^{0.33}$.  Thus our
numerical treatment suggests that no new scattering rate combinations
are produced in the Jones-Zener expansion until the magnetoresistance
saturates. We have confirmed this is the case analytically at fourth
order in the magnetic field.
This scaling also means that in our model there is an absolute value of
magnetoresistance above which one must see a deviation from a $B^2$
dependence.  Our model predicts that with a magnetoresistance of
only 0.5\% one would expect at least a 10\% deviation from a $B^2$
form.  Comparing this with experimental work~\cite{tyler_1998a} on
optimally doped Tl-2201 in pulsed magnets, we find a deviation from
quadratic dependence is only perhaps being seen in optimally doped
Tl-2201 at a magnetoresistances of order 3\%. 
(The observed deviation from the weak-field region is more consistent
with the Coleman-Schofield-Tsvelik
phenomenology which predicts that this should occur when
the magnetoresistance is at or less
than about 10\%~\cite{tyler_1998a}.)

\subsection{$c$-axis transport}

In-plane magnetoresistance in the cold-spot model is at odds with the
current experiments. It has been argued that including vertex
corrections could correct this~\cite{narikiyo_2000a,kontani_1999a}.
As we have argued in the introduction, $c$-axis transport is then of
fundamental importance in testing the cold-spot model. This is because
to lowest order in $t_\perp^2$ the $c$-axis conductivity is related to
the convolution of two in-plane spectral
functions~\cite{mckenzie_1999a}. 
These are the
quantities which may be inferred from the angle-resolved photoemission
that inspired the cold-spot model. Thus, independent of a picture of
quasiparticles for in-plane transport, a cold-spot model should be
able to account for the $c$-axis d.c. conductivity~\cite{millis_2000a}.
In particular we must show that this leads to a consistent picture of
$c$-axis orbital magnetoresistance (magnetic field in the plane and
electric field out of the plane). The most systematic study of these
effects has been performed by Hussey {\it et al.}~\cite{hussey_1996a}
in overdoped Tl-2201 and this motivates the following analysis. The
two-lifetime behavior is much less apparent in the overdoped material
so we will not be able to use the $\Gamma_f \gg \Gamma_s$
asymptotics. Instead we need a full solution for any degree of
anisotropy. 
  
We will now calculate the $c$-axis magnetoresistance,
$\Delta\rho_{zz}/\rho_{zz}$.  To obtain this, we first calculate the
relevant components of the conductivity matrix, ${\sigma_{ij}}$ and
invert.  In this experimental geometry, $\sigma$ has zero
components $\sigma_{xy}$,$\sigma_{xz}$,$\sigma_{yx}$ and $\sigma_{zx}$
due to the ${\mathbf B}$ field being in-plane.  $\sigma_{xx}$ is equal
to the zero-magnetic field value of $\sigma_{yy}$.  We can expand the
other terms in magnetic field:
%
%
\begin{equation}
\sigma_{\nu\mu}=\sigma_{\nu\mu}^{(0)} 
+ \sum_{n=1}^{n=3}{\sigma_{\nu\mu}^{(n)}},
%
\label{sigmaexpan2}  
\end{equation}
where the superscripts refer to the order of effect in magnetic field.
Symmetry under time reversal means, of course, that 
$\sigma_{zz}^{(1)}$ and $\sigma_{zz}^{(3)}$ and the Hall terms
$\sigma_{yz}^{(0)}$ and $\sigma_{yz}^{(2)}$ are zero. In addition
the Hall term $\sigma_{yz}^{(1)}$ is small ($\sim t_{\perp}^2$).  Under these
conditions, the $c$-axis magnetoresistivity in weak magnetic field
simplifies to
\begin{eqnarray}  
\rho_{zz}&=&\rho^{(0)}_{zz} + \Delta\rho_{zz}^{(2)}
-\Delta\rho_{zz}^{(4)} +O[B^6]\; , \nonumber \\
{\Delta\rho_{zz}^{(2)}\over\rho^{(0)}_{zz}}&=&-{{\sigma_{zz}^{(2)}}
\over {\sigma_{zz}^{(0)}}}\; , \nonumber \\
{\Delta\rho_{zz}^{(4)}\over\rho^{(0)}_{zz}}&=&
-{{(\sigma_{zz}^{(2)}-\sigma_{zz}^{(0)}\sigma_{zz}^{(4)})}\over
{\sigma_{zz}^{(0)\, 2}}} \;.
\label{rhoexpan} 
\end{eqnarray} 
Here we have adopted the sign convention of Hussey {\it et
al.}~\cite{hussey_1996a} and Dr\v{a}gulescu {\it et
al.}~\cite{dragulescu_1999a}, expecting the fourth-order
magnetoresistivity to be negative.

To obtain the conductivity, we follow the same general procedure as
for the in-plane case, solving the Boltzmann equation using a
Jones-Zener expansion. We now introduce the angle, $\phi$, which
is the in-plane angle between the B field and the $a$-direction.
$\theta$ becomes the azimuthal coordinate relative to the direction of
the B field. The solutions to the Boltzmann equation are now
\begin{equation}
\psi({\mathbf k})=
{e E \over \hbar\Gamma(\theta+\phi)}{\partial\epsilon \over \partial
k_{\perp}}+\sum_{n=1}^\infty \psi^{(n)}({\mathbf k}) \; ,
\label{Zenerc0}
\end{equation}
where, to lowest order in $t_{\perp}$, 
\begin{equation}
\psi^{(n)}({\mathbf k})=
-{eB^{n} \over \hbar\Gamma(\theta+\phi)}
\left(
{1\over\hbar}v_F\sin\theta
{\partial\psi^{(n-1)} \over \partial k_{\perp}}\right) \, .
\label{Zenercnsimple}
\end{equation}
The ratio of $\sigma_{zz}^{(2)}/\sigma_{zz}^{(0)}$, when
expanded in powers of $\alpha$ yields a zero-order term which is, as Hussey
{\it et al.}~\cite{hussey_1996a} first showed 
\begin{equation}
{\sigma_{zz}^{(2)}\over \sigma_{zz}^{(0)}}=
- {{c^2\,e^2\,v_{F}^2} \over {2\hbar^2 \Gamma_{0}^{2}}}
+ O[\alpha^2] \; .
\label{sigmacresult2}  
\end{equation}
Similarly, for the quadratic term in the $c$-axis conductivity:
\begin{equation}  
{\sigma_{zz}^{(4)}\over \sigma_{zz}^{(0)}}/\Bigl({\sigma_{zz}^{(2)}\over
\sigma_{zz}^{(0)}}\Bigr)^2=
{6 + 3\gamma^{2} +  2\gamma\cos (4\,\phi ) \over 
  2(2 + \gamma^{2}) } \; ,
\label{sigmacresult4}  
\end{equation}
which in the limit of simple inter-plane tunneling ($\gamma \rightarrow
0$) gives 3/2, again as shown by Hussey {\it et al.}
%

The Hall conductivity, $\sigma_{yz}$, is found to vary
as $t_{\perp}^2$. In the limit of isotropic in-plane scattering the
out-of-plane Hall angle is identical to the in-plane
one~\cite{schofield_2000a} 
\begin{equation}  
{\text{lim}}_{\alpha\rightarrow 0} {\sigma_{yz}^{(1)} \over
\sigma_{zz}^{(0)}} = {eBv_F \over \hbar k_F\Gamma_{0}} \; .
\label{cHallterm}  
\end{equation}
This is not true in general however. Nevertheless its dependence on
$t_\perp^2$ means that the Hall conductivity does not appear, at
leading order, in the out-of-plane magnetoresistance.

\begin{figure}
\includegraphics[width=\columnwidth]{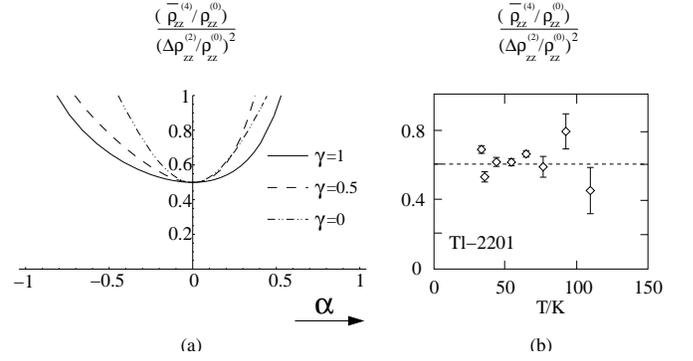}
\caption{(a) $\bar{\rho}_{zz}^{(4)}/\rho^{(0)}_{zz} \over
(\Delta\rho^{(2)}_{zz}
/\rho^{(0)}_{zz})^2$ and (b) experimental data~\cite{hussey_1996a}
from measurements on Tl$_2$Ba$_2$CuO$_6$, where this quantity was
$0.6\pm0.1$, approximately independent of temperature.  This implies that 
$\alpha$ should not vary greatly in the experimental range of
temperatures.}
\label{figure5}
\end{figure}

Rather than display the full expression for every quantity we have
calculated, we show graphically two quantities which Hussey {\it et
al.} examined experimentally in single-crystal Tl$_2$Ba$_2$CuO$_6$.
There a key observation was the four-fold variation of the $c$-axis
magnetoresistance as the magnetic field was rotated in-plane, {\it
i.e.} that
\begin{equation}  
\Delta\rho_{zz}^{(4)}= 
\bar{\rho}_{zz}^{(4)} + \tilde{\rho}_{zz}^{(4)}\cos(4\phi) \, ,
\label{rho4expan}  
\end{equation}
where $\bar{\rho}_{zz}^{(4)}$ and $\tilde{\rho}_{zz}^{(4)}$ were both
found to be positive.  In fact it is straightforward to show that this
4-fold modulation at fourth order in B is a simple consequence of
square symmetry.  Hussey {\it et al.} initially analyzed their
angle-dependent magnetoresistance results purely in terms of
anisotropic, in-plane scattering (as parameterized here by $\alpha$).
We first look at the offset part in the fourth order magnetic field
term in the magnetoresistance.  We examine
$\bar{\rho}_{zz}^{(4)}/\rho^{(0)}_{zz} \over
(\Delta\rho^{(2)}/\rho^{(0)}_{zz})^2$, a quantity found experimentally
to be roughly independent of temperature and approximately equal to
0.6 $(0.6\pm0.1)$.  We show it for various values of $\gamma$, as a
function of $\alpha$ in Figure \ref{figure5}.  This puts a
constraint on the temperature dependence of $\alpha$, an issue which
is explored in section~\ref{application} where we try to fit
magnetoresistance data to our model.
%
%

\begin{figure}
\includegraphics[width=\columnwidth]{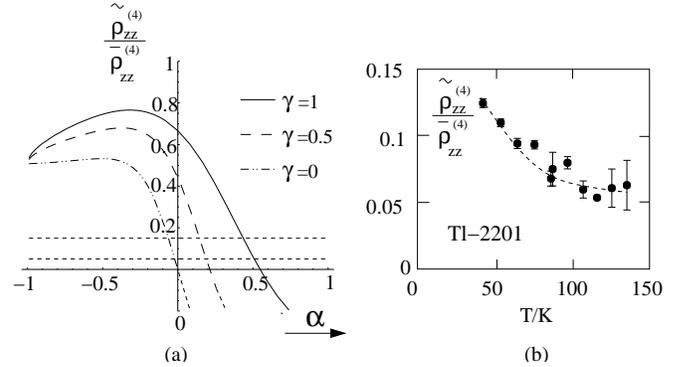} 
\caption{(a) The size of the fourth order modulation in c-axis
magnetoresistivity, relative to the offset at this order and (b) 
experimental data~\cite{hussey_1996a} from measurements on
Tl$_2$Ba$_2$CuO$_6$ again shown for comparison.  A positive value of
$\tilde{\rho}_{zz}^{(4)}/\bar{\rho}_{zz}^{(4)}$ means that resistance
maxima exist for a magnetic field along [110].  The effect of
bandstructure is therefore pronounced and produces 
resistance maxima where the cold spots are on the zone diagonals ($\alpha
>0$).  In (a), the approximate constraints on the valid regions of
$(\alpha,\gamma)$ parameter space are given by the limits of the
modulation as taken from experiment.  These are shown as dotted lines at
values of 0.15 and 0.05.}
\label{figure6}
\end{figure}

The second quantity studied experimentally is the amplitude of the
modulation in magnetoresistance at order $B^4$.  The quantity
evaluated is $\tilde{\rho}_{zz}^{(4)}/\bar{\rho}_{zz}^{(4)}$, which is
positive when the resistance is maximum with ${\mathbf B}$ aligned
along the zone diagonal (Eq. \ref{rho4expan}). This too is plotted
across the range of scattering anisotropy, $\alpha$, for various
bandstructures, $\gamma$, in Figure \ref{figure6}. 

Hussey {\it et al.} found it to vary with temperature, dropping from
about 0.15 at 50K to 0.05 at about 140K, and these limits are shown as
dotted lines in Figure \ref{figure6}.  In our calculations,
$\tilde{\rho}_{zz}^{(4)}/\bar{\rho}_{zz}^{(4)}$ is positive when
$\alpha$ is negative, corresponding to a situation where the hot spots
and cold spots have effectively exchanged their usual positions.  This
can be explained intuitively by identifying the maxima in the Lorentz
force, ${\mathbf v} \times {\mathbf B}$, at points on the Fermi surface
orthogonal to the in-plane magnetic field.  Then, loosely, the parts of
the Fermi surface least affected by a magnetic field are those with
${\mathbf v_{F}}$ parallel to ${\mathbf B}$. Thus, naively, one would
expect the {\em minima} in the magnetoresistance to occur when the field
is along the cold-spot directions. In fact experimentally this
orientation gives a maxima in the magnetoresistance. Turning on
anisotropy in the $c$-axis dispersion $\gamma$, one can sufficiently
inhibit c-axis transport through the cold spots in the {\it positive}
$\alpha$ regime that resistance maxima can be seen for ${\mathbf B}$
along [110] with cold spots along [110] in keeping with the experiments.

This effect has been attributed elsewhere~\cite{dragulescu_1999a} to
the curvature of the real Fermi surface in a model of isotropic
scattering.  Here we have shown that the two effects will work
together in the regime of positive bandstructure anisotropy.
Once again, however, we see that there is a constraint on the allowed
degree of variation of scattering anisotropy with temperature.
In Table~\ref{rhoctable} we show the form of these 4th order quantities
shown in Figures \ref{figure5} and \ref{figure6} as well as the size
of the 2nd order magnetoresistivity and in-plane properties.  This will be
useful when fitting to data later. 
%
\begin{table*}
\begin{tabular}{|c|c|c|c|c|c|c|c|}
\hline
Quantity 
& $\rho_{xx}^{(0)}$ 
& $\cot\Theta_{H}$
& ${\Delta\rho_{xx}^{(2)} \over \rho_{xx}^{(0)}}$
& ${\Delta\rho_{xx}^{(2)} \rho_{xx}^{(0)}}$
& ${\Delta\rho_{zz}^{(2)} \over \rho_{zz}^{(0)}}$
& $(\bar{\rho}_{zz}^{(4)} /\rho_{zz}^{(0)}) \over (\Delta\rho_{zz}^{(2)}
/\rho_{zz}^{(0)})^2$
& $\tilde{\rho}_{zz}^{(4)}/\bar{\rho}_{zz}^{(4)}$\\
&&&&&&&\\
Cold-spot &
${2\pi\hbar c \Gamma_{0} \over e^2 v_F k_F}(1-\alpha^2)^{1/2}$ 
& ${\hbar k_{F}\Gamma_{0}(1-\alpha^2) \over eBv_{F}}$
& $({eBv_F \over \hbar k_{F} \Gamma_{0}})^2{\alpha^2(17+3\alpha^2) \over
2(1-\alpha^2)^3}$ 
& $({2\pi cB \over k_{F}^{2}e})^2 {\alpha^2(17+3\alpha^2) \over
2(1-\alpha^2)^2}$
& $({cev_{F}B \over \hbar \Gamma_{0}})^2 f_{1}(\alpha,\gamma)$
& $ f_{2}(\alpha,\gamma)$
& $ f_{3}(\alpha,\gamma)$ \\
model &&&&&&&\\
\hline
\end{tabular} 
\smallskip 
\caption{Results of our calculations of in-plane and
$c$-axis properties. In particular we have attempted to find
dimensionless quantities, $f_{1}$, $f_{2}$ and $f_{3}$, which are
functions of $\alpha$ and $\gamma$. This enables us to constrain the
values of $\alpha$ and $\gamma$ by comparing directly to experiments
without further assumptions.}
\label{rhoctable} 
\end{table*} 

%
\subsection{Non-ohmic conductivity} 
We have also calculated the non-ohmic, in-plane conductivity of
the system in zero magnetic
field.  Hlubina~\cite{hlubina_1998a} has suggested that this is a
useful test of the 
model of Ioffe and Millis.  The Boltzmann equation
(Eq.~\ref{simpleBoltz}) is now solved by
Jones-Zener expansion in ${\mathbf{E}}$ and is given by 
\begin{equation}
\Gamma_{{\mathbf k}}\psi- {e \over\hbar} {\mathbf E}\cdot{\nabla_{{\mathbf
k}}\psi}=e{\mathbf E}\cdot{\mathbf v_{\mathbf k}}. 
\label{simplenonlinBoltz}
\end{equation} 
From the 4-fold symmetry of the system, it is straightforward
to deduce that the currents in the $x$ and $y$ directions must have 
the following form at third order in the electric field:
\begin{eqnarray}  
j_{x}&=&\sigma_{0}E_{x}+\sigma_{1}E_{x}^{3}+\sigma_{2} E_{y}^{2} E_{x}
\; ,\\
j_{y}&=&\sigma_{0}E_{y}+\sigma_{1}E_{y}^{3}+\sigma_{2} E_{x}^{2} E_{y}
\; .
\label{nonohmj}
\end{eqnarray}
We then find that 
\begin{eqnarray}  
\sigma_{0}&=&{e^2 k_{F}v_{F} \over 2\pi\hbar c
\Gamma_{0}(1-\alpha^2)^{1/2}} \; , 
\label{nonohm0} \\
\sigma_{1}&=&-{e^4 v_{F} \over 4\pi\hbar^3 k_{F} c \Gamma_{0}^{3}}
{\alpha((1+\alpha)(\alpha^2+2\alpha+2)) \over
(1-\alpha^2)^{7/2}} \; , 
\label{nonohm1} \\
\sigma_{2}&=&-{e^4 v_{F} \over 4\pi\hbar^3 k_{F} c \Gamma_{0}^{3}}
{\alpha((\alpha^3-9\alpha^2+4\alpha-6) \over
(1-\alpha^2)^{7/2}} \; ,
\label{nonohm2}  
\end{eqnarray}
which all simplify or reduce to zero appropriately in the isotropic limit,
$\alpha \rightarrow 0$.  Furthermore, in the limit of strong
anisotropy, $\alpha\rightarrow 1$, we find
$\sigma_{1}=-\sigma_{2}$, 
as demonstrated by Hlubina~\cite{hlubina_1998a}.

The most obvious consequence of non-linear response in the electric field
is that the current no longer flows parallel to the electric
field. Defining the parallel current ($j_{||}$) as being the response
parallel to the applied field, and the transverse current as the
current component perpendicular to the applied field ($j_\perp$) we
find that 
\begin{equation}
j_\perp = {1 \over 4} E^3 \left( \sigma_1 - \sigma_2 \right) \sin 4
\phi \; , 
\end{equation}
where $\phi$ is the angle between the in-plane electric field and the
$a$-axis. Using our expressions for $\sigma_1$ and $\sigma_2$
(Eqs.~\ref{nonohm1} and~\ref{nonohm2}) we may
write
\begin{eqnarray}
{j_\perp \over j_{||}} &=& \left({e E \over \hbar k_F \Gamma_0}\right)^2 {
\alpha \left(2 + \alpha^3 \right) \over 2 \left(1- \alpha^2 \right)^3}
\sin 4\phi \; , \\
&=&\lim_{\alpha\rightarrow 1} \left({v_d \over
v_F}\right)^2 {5\over 16} \left({\Gamma_f \over \Gamma_s}\right)^3 
\sin 4 \phi\; . 
\end{eqnarray}
Here $v_d= e E/2 m\Gamma_0$ is the Drude drift velocity of a
fast decaying quasiparticle. So, although the $v_d/v_F$ is generally tiny,
nonlinear effects are indeed strongly enhanced by anisotropy.

\subsection{Thermal transport}
In this treatment we have not speculated on the anisotropy of the
energy relaxation rate around the Fermi surface. However if we make the
assumption that the energy relaxation follows the quasiparticle
relaxation rate and only varies around the Fermi surface (as opposed to
variations away from the Fermi surface) we can calculate all of the
thermal transport coefficients. Without reproducing the details of the
calculation we can see immediately from the form of the Boltzmann
equation (Eq.~\ref{simpleBoltz}) that temperature gradients drive the
quasiparticle distribution in exactly the same way as electric fields
except for the usual factor of $(\epsilon_{\mathbf k} -
\mu_F)/T$. This is seen in the right-hand-side of
Eq.~\ref{simpleBoltz}. This factor is isotropic around the Fermi
surface and does not introduce any further angular dependence. Thus we
can conclude that all currents proportional to temperature gradients
will have exactly the same dependence on scattering rate as those
driven by electric fields (for example $\sqrt{\Gamma_f\Gamma_s}$ in the
absence of a magnetic field). Within this approximation the
thermal conductivity will obey the Wiedemann-Franz law and the
diffusion thermopower will have the usual linear temperature
dependence and be independent of the scattering rate. To account for
the unusual systematics in the measured thermopower of the cuprate
metals~\cite{obertelli_1992a} in this model one would need to add a
significant anisotropic energy dependence of the scattering rate that
differs from the anisotropy of the transport relaxation rate.

%
\section{Application to experimental data}
\label{application}
In order to apply the analytic results of our model to experimental
data, we have made a comparison of magnetoresistivity results with
measurements on overdoped samples of single-crystal
Tl$_2$Ba$_2$CuO$_6$, made by Tyler~\cite{tyler_1998b} and
Hussey~\cite{hussey_1996a}. We choose this system because its $c$-axis
magnetoresistance is well characterized. We have argued that $c$-axis
properties are the most robust quantities in this model since they are
not affected by vertex corrections, nor do they rely on a
quasiparticle picture. We then identify combinations of measured
$c$-axis quantities which directly probe the degree of anisotropy with
minimal dependence on unknown parameters. Finally we address the
in-plane transport features.
We also identify
combinations of experimental quantities which allow a comparison with
theory involving the fewest assumptions on unknown parameters. In
particular we can constrain the degree of anisotropy in the $c$-axis
dispersion ($\gamma$) and the range of scattering anisotropy
($\alpha$) from the $c$-axis magnetoresistance.  

First we consider the overall magnitude of the fourth order
magnetoresistance by comparing this to the second order term. This is
shown in Fig.~\ref{figure5}(b). When we compare this to our model
[Fig.~\ref{figure5}(a)] we see that to obtain the observation of
an essentially temperature independent result of around 0.6, the degree
of scattering anisotropy must be limited to $\alpha <
0.5$. Furthermore, if $\alpha$ is to have any temperature dependence
at all then $\gamma$ should be approaching 1 where the gradients in
Fig.~\ref{figure5}(a) are smallest.

We have already remarked on a second conclusion from a
comparison between theory and experiment. In order for the maxima in
the $c$-axis magnetoresistance to occur when the field is along the
zone-diagonals, $c$-axis transport along these directions must be
suppressed. This is illustrated in Fig.~\ref{figure6}(a) where we
see that with $\alpha >0$ we require $\gamma>0$ to obtain a ratio of
$\tilde{\rho}_{zz}^{(4)}/\bar{\rho}_{zz}^{(4)}$ of the correct sign.
Furthermore the experimental bounds on this ratio [illustrated in
Fig.~\ref{figure6}(b)] confirm that the degree of scattering
anisotropy ($\alpha$) must be less than about 0.6. In no sense then
are we in the limit of strong anisotropy in this overdoped
material. 

A further observation comes from the temperature dependence shown in
Fig.~\ref{figure6}(b). Generically the quantities shown in
Figs.~\ref{figure5}(b) and~\ref{figure6}(b) should be
temperature dependent so the observation that only the second of these
has any significant variation with temperature restricts $\alpha$ to
vary between about 0.2 and 0.4 with a band structure anisotropy
$\gamma \sim 0.9$. Also, in order to fit the trends in the data, we
see that the overdoped state must become less anisotropic as the
temperature is lowered. Proximity to a zero temperature critical point
would suggest the opposite should be true but perhaps here elastic
processes are begining to dominate. 

In order to predict other experimental quantities, we then need to
know the temperature dependence of $\Gamma_{0}$.  This too can be
found from the second order c-axis magnetoresistance, when data is
compared with the functional form in table \ref{rhoctable}, here
fixing $\gamma=1$. In addition it can also be independently infered
from the in-plane resistivity but now relying on the assumptions of
Boltzman transport theory.

\begin{figure}
\includegraphics[width=\columnwidth]{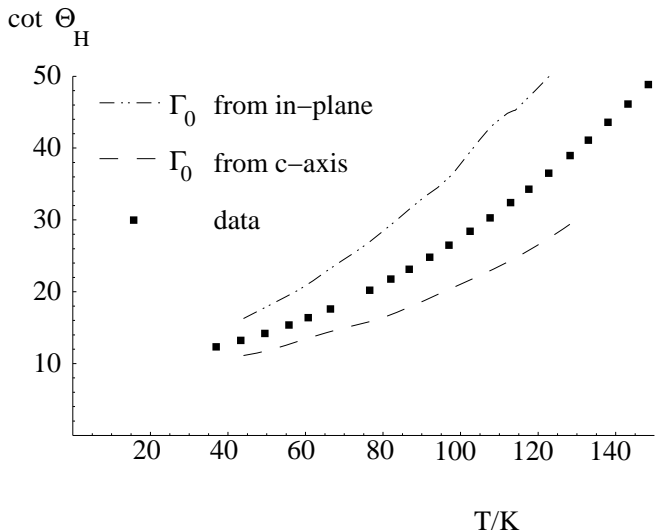} 
\caption{A comparison of our theoretical prediction for
$\cot\Theta_{H}$ with experimental data~\cite{tyler_1998b} from
measurements on Tl$_2$Ba$_2$CuO$_6$.  Predictions using $\Gamma_{0}$
found from both in-plane and out-of-plane properties are shown, relating
to using the transport or the intrinsic quasiparticle lifetime.}
\label{figure7}
\end{figure}

\begin{figure}
\includegraphics[width=\columnwidth]{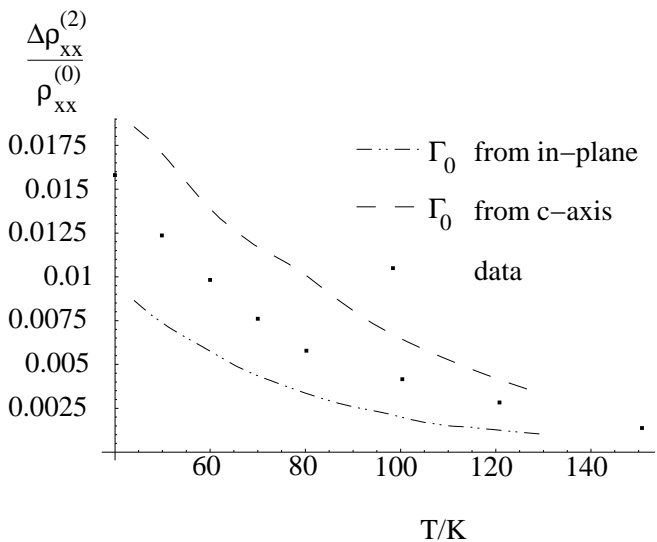} 
\caption{A comparison of our theoretical prediction for
in-plane, second order magnetoresistance with experimental
data~\cite{tyler_1998b} from measurements on Tl$_2$Ba$_2$CuO$_6$.
Again, predictions using $\Gamma_{0}$ found from both in-plane and
out-of-plane properties are shown.}
\label{figure8}
\end{figure}

With these two temperature dependencies known, we may try to predict
in-plane properties.  We show plots of the cotangent of the Hall
angle, $\cot\Theta_H$, and the in-plane magnetoresistance
${\Delta\rho_{xx}^{(2)}/\rho_{xx}^{(0)}}$, together with experimental
data from Hussey and Tyler in Figs.~\ref{figure7}
and~\ref{figure8}. We see by comparing the experimental data
with the parameters extracted from other measurements that the data
and theory follow the same trends and are of the correct order of
magnitudes. The two different fits on each plot represent the two
different methods of infering $\Gamma_0$: either using the in-plane
resistivity or the out-of-plane magnetoresistance.
The difference in these two
methods is that the in-plane measurements would be expected to be
sensitive to the transport lifetime, as opposed to the quasiparticle
lifetime in the case of out-of-plane measurements. Surprisingly it 
appears that the quasiparticle lifetime is longer than the transport
lifetime. 


Overall we see that for overdoped Tl-2201 a consistent picture
emerges
of a Fermi-liquid-like metal with weak scattering anisotropy around
the Fermi surface but a strong anisotropy in the $c$-axis
dispersion. A more rigourous test of the model would be obtained by
performing a similar comparison on optimally doped Tl-2201 ($T_c
=85$K). There the two-lifetime behavior of in-plane magnetotransport
is clearly apparent with very good fits to data being found with
simple power laws~\cite{tyler_1998b}:
\begin{eqnarray}
\rho_{xx} &=& -20 + 1.56 (T/{\rm K}) \; \mu\Omega{\rm cm} \; , \\
(B/{\rm T}){\cot\theta_H} &=& 300 + 1.78*10^{-2} (T/{\rm K})^2 \; , 
\end{eqnarray}
for $100{\rm K} < T < 300$K.
In the limit of high in-plane
anisotropy we can  combine these measurements to obtain the
magnetoresistance. We find
\begin{equation}
{\Delta \rho_{xx} \over \rho_{xx}^{(0)}} = 
{5 \over 2} \left( e^2 k_F^4 \over \pi^2 c^2 B^2 \right) \left[{
\rho_{xx}^{(0)} \over \cot^2 \theta_H} \right]^2
\; . 
\end{equation}
This would predict a magnetoresistance of about 0.04 at 10T and 130K
(so already out of the weak-field regime)---a factor of 40 greater
than currently seen. We have used in-plane properties to predict the
magnetoresistance so uncertainties about vertex corrections are
present in this estimate.

Again, using the in-plane properties, we would expect that the
out-of-plane magnetoresistance would be 0.1 in fields as low as 3T
with a fourth order magnetoresistance of 0.001 in an in-plane field of
about 5T. These effects should be measurable but again rely on using
in-plane transport lifetimes to provide a measure of in-plane
quasiparticle lifetimes. However, we can make an unambiguous
prediction within this model: that the positions of the maxima in
resistivity at fourth order should move from the [110] directions
(seen in the overdoped system) to the [100] directions in the
optimally doped materials. This is seen in Fig.~\ref{figure6}(a)
where for large $\alpha$ we see that bandstructure is unable to
prevent the sign of the modulation term,
$\tilde{\rho}_{zz}^{(4)}/\bar{\rho}_{zz}^{(4)}$, from being negative.


\section{Conclusion}
\label{Conclusion}

We have presented a thorough investigation of a minimal model for
transport in a quasi-2D system, where we allow for anisotropic
scattering in-plane and an anisotropy in the out-of-plane
dispersion, both fully variable.  Such a model has been proposed for
the normal state of the cuprate superconductors with either strong
scattering hot spots and weak scattering cold regions around the
Fermi surface or hot regions with cold spots. We have studied the
transport properties of both types of models and illustrated how
short-circuiting in the hot-spot model makes this inconsistent with
transport measurements on the cuprates. For the cold-spot model we
have computed in- and out-of-plane magnetoresistivity, in-plane
non-ohmic conductivity and thermal conductivities for an arbitrary
degree of anisotropy.

We find that the in-plane magnetoresistance in this model is too large
when compared with experiments on the cuprates, in keeping with other
work~\cite{ioffe_1998a}. In addition the in-plane magnetoresistance
should be universal beyond the weak field limit with a well defined
deviation from a $B^2$ dependence at weaker fields than currently
observed.  It has been argued that vertex corrections may account for
these discrepancies between the model and experiment.  Here we have
focussed on the out-of-plane magnetoresistance for which there are no
vertex corrections in the quasi-2D limit. We have completely
characterized the magnetoresistance to fourth order in this
geometry. We have compared our model with experiments on overdoped
Tl2201 and show that the model can be reasonably well fit to the
experiments. This requires only weak scattering anisotropy but a high degree
of bandstructure anisotropy in the $c$-axis dispersion.

A better test of this model would be to compare it to optimally doped
Tl-2201 where the two-lifetime behavior is very clearly seen in
in-plane magnetotransport measurements. This model would predict that
strong angle dependent $c$-axis magnetoresistance should be observed
and the positions of the minima in the optimally doped system should
be rotated by $\pi/4$ from those of the overdoped material. A
quantitative analysis of the degree of in-plane anisotropy could also
be made. The extent to which vertex corrections control the in-plane
transport properties could then be assessed.

\section*{Acknowledgements} The authors would like to thank C. Bergemann,
D. R. Broun, N. E. Hussey, S. R. Julian, D. E. Khmelnitskii,
M. W. Long, G. G.  Lonzarich, A. J. Millis, A. Rosch and T. Xiang for
useful discussions.  We are grateful for the hospitality of the
Center for Materials Theory, Rutgers University and the Newton
Institute, Cambridge where some of this work was done and also for
the support of the Theory of Condensed Matter Group in
Cambridge.  KGS was funded by the EPSRC, AJS by the Royal Society.

\end{document}